\documentclass[10pt,letterpaper]{article}
\usepackage{opex3}
\usepackage{graphicx}
\usepackage{bm}

\begin{document}

\title{3D phase-matching conditions for the generation of entangled
triplets by $\chi^{(2)}$ interlinked interactions}

\author{Maria Bondani}

\address{
National Laboratory for Ultrafast and Ultraintense Optical Science,
Consiglio Nazionale delle Ricerche, Istituto Nazionale per la Fisica
della Materia, Unit\`a di Como, via Valleggio 11 - 22100 Como,
Italy}
\email{maria.bondani@uninsubria.it} 

\author{Alessia Allevi, Eleonora Gevinti, Andrea Agliati, Alessandra Andreoni}

\address{
Dipartimento di Fisica e Matematica, Universit\`a degli Studi
dell'Insubria and Istituto Nazionale per la Fisica della Materia,
Unit\`a di Como, via Valleggio, 11 - 22100 Como, Italy}



\begin{abstract}
An analytical calculation of the interaction geometry of two
interlinked second-order nonlinear processes fulfilling
phase-matching conditions is presented. The method is developed for
type-I uniaxial crystals and gives the positions on a screen beyond
the crystal of the entangled triplets generated by the interactions.
The analytical results are compared to experiments realized in the
macroscopic regime. Preliminary tests to identify the triplets are
also performed based on intensity correlations.
\end{abstract}

\ocis{(190.4410) Nonlinear optics, parametric processes; (270.0270)
Quantum optics.}


\section{Introduction}
The production of multipartite entangled states by means of multiple
nonlinear interactions occurring in a single nonlinear crystal has
been recently suggested \cite{Ferraro2004,Rodionov2004,Bradley2005}
as an alternative to the use of single-mode squeezed states
\cite{vanLoock2000,Jing2003,Aoki2003} or of two-mode entangled
states and linear optical elements \cite{Glockl2003}. As in the case
of bipartite states produced by spontaneous parametric
downconversion, such multipartite states display entanglement also
in the macroscopic regime in which bright outputs are generated: for
this reason they are particularly interesting for all the
applications of continuous-variable entanglement
\cite{QI2003,Braunstein2005}.\\
The possibility of realizing the simultaneous phase-matching (PM) of
two traveling-wave parametric processes in a single crystal in a
seeded configuration has been already demonstrated
\cite{Lee2003,Bondani2004}. In this paper we present the
experimental realization of the same interlinked interactions
starting from vacuum fluctuations. The output of the nonlinear
crystal displays the entire ensemble of inseparable tripartite
entangled states satisfying the PM conditions. In order to identify
a triplet, we develop a 3D calculation of the characteristics of
interlinked PM that reproduce the experimental phenomenology.
Finally we perform a preliminary verification of the number of
photons conservation law by means of intensity correlations.
\section{Theory}
We consider the Hamiltonian describing the simultaneous PM of a
downconversion and an upconversion processes
\begin{equation}  \label{intH2} H_{\mathrm{int}} = \gamma_1
a_1^\dag a^{\dag}_3 + \gamma_2 a_2^{\dag} a_3 + h.c.\;,
\end{equation}
where $\gamma_1\propto a_4$ and $\gamma_2\propto a_5$ are coupling
coefficients.  The two interlinked interactions involve five fields
$a_{\mathrm j}$, two of which, say $a_4$ and $a_5$, enter the
crystal and act as non-evolving pumps. When acting on the vacuum as
the initial state, $H_{\mathrm{int}}$ admits the following
conservation law
\begin{equation}
N_1(t) = N_2(t) + N_3(t) \label{cm}\;,
\end{equation}
being $N_{\mathrm j}(t)=\langle a^\dag_{\mathrm j}(t) a_{\mathrm
j}(t)\rangle$ the mean number of photons in the j-th mode, and
yields a fully inseparable tripartite state \cite{Ferraro2004}.
We consider the realization of Eq.~(\ref{intH2}) in a negative
uniaxial crystal in type-I non-collinear PM interaction geometry.
The processes must satisfy energy-matching
($\omega_{4}=\omega_{1}+\omega_{3}$,
$\omega_{2}=\omega_{3}+\omega_{5}$) and PM conditions (${\mathbf
k^e_4}={\mathbf k^o_1}+{\mathbf k^o_3}$, ${\mathbf k^e_2}={\mathbf
k^o_3}+{\mathbf k^o_5}$), where $\omega_{\mathrm{j}}$ are the
angular frequencies, ${\mathbf k_{\mathrm{j}}}$ are the wavevectors
and $^{o,e}$ indicate ordinary and extraordinary field
polarizations. In order to investigate the geometrical constraints
imposed by the PM conditions, we analytically calculate the output
angles of each interacting field.
\begin{figure}[h]
\centering
\includegraphics[width=0.4\textwidth,angle=270]{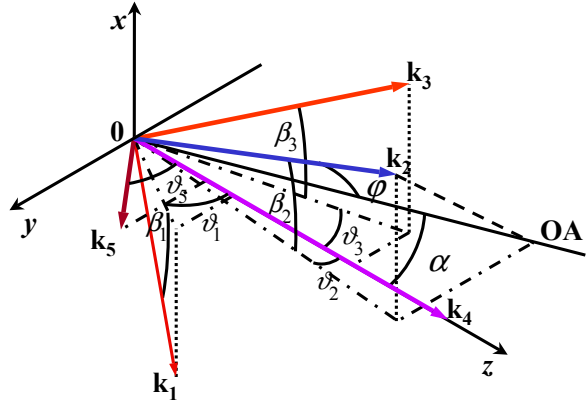}
\caption{\label{f:inter} Scheme of the phase-matched interlinked
interactions: $(x,y)$-plane coinciding with the crystal entrance
face; $\alpha$, tuning angle; $\beta_{\mathrm j}$'s, angles to
$(y,z)$-plane; $\vartheta_{\mathrm j}$'s, angles on the
$(y,z)$-plane; $\varphi$, angle to the optical axis (OA).}
\end{figure}
\par
To simplify calculations, we assume, in the reference frame depicted
in Fig.~\ref{f:inter}, that the two pumps lie in the $(y,z)$-plane
containing the optical axis (OA) and the normal to the crystal
entrance face and that the pump field $a_4$ propagates along the
normal, $z$. The solutions corresponding to the effective
experimental orientation of the crystal can be obtained by simply
calculating the refraction of the beams at the crystal
entrance/output faces. Accordingly, the PM conditions for the two
interactions simultaneously phase-matched can be written as
\begin{eqnarray}
k_1 \sin\beta_1 + k_3 \sin\beta_3 &=& 0\label{PM1a}\\
k_1 \cos\beta_1 \sin\vartheta_1 + k_3 \cos\beta_3 \sin\vartheta_3 &=& 0\label{PM1b}\\
k_1 \cos\beta_1 \cos\vartheta_1 + k_3 \cos\beta_3 \cos\vartheta_3
&=& k_4\label{PM1c}\\
k_2 \sin\beta_2 &=& k_3 \sin\beta_3\label{PM2a}\\
k_2 \cos\beta_2 \sin\vartheta_2 &=& k_3 \cos\beta_3 \sin\vartheta_3 + k_5 \sin\vartheta_5\label{PM2b}\\
k_2 \cos\beta_2 \cos\vartheta_2 &=& k_3 \cos\beta_3 \cos\vartheta_3
+ k_5 \cos\vartheta_5\label{PM2c}
\end{eqnarray}
where the angles are defined as in Fig.~\ref{f:inter}. The
wavevectors $k_{\mathrm j}$ are defined as $k_{\mathrm j}=
n_{\mathrm j}(\omega_{\mathrm
j},{\mathbf{\hat{k}_j}})\omega_{\mathrm j}/c$, being $c$ the speed
of light in the vacuum and $n_{\mathrm j}(\omega_{\mathrm
j},{\mathbf {\hat{k}_j}})$ the refraction indices of the medium
\begin{eqnarray}
n_j\left(\omega_j\right) = n_o\left(\omega_j\right)&&\quad j=1,3,5\\
n_2\left(\omega_2,\varphi\right)=
\left[\frac{\cos^2\varphi}{n_o^2(\omega_2)}+
\frac{\sin^2\varphi}{n_e^2(\omega_2)}\right]^{-1/2} &;&
n_4\left(\omega_4,\alpha\right)=
\left[\frac{\cos^2\alpha}{n_o^2(\omega_4)}+
\frac{\sin^2\alpha}{n_e^2(\omega_4)}\right]^{-1/2}\label{index}\;.
\end{eqnarray}
where $n_{o,e}(\omega)$ are given by the dispersion relations of the
medium. Note that $\cos\varphi=\cos\beta_2\cos(\vartheta_2-\alpha)$
and $\beta_5=\beta_4=\vartheta_4=0$.
\par
For fixed frequencies and propagation directions of the pump fields,
we are left with 11 variables ($\omega_1$, $\omega_2$, $\omega_3$,
$\vartheta_1$, $\vartheta_2$, $\vartheta_3$, $\vartheta_5$,
$\beta_1$, $\beta_2$, $\beta_3$, $\alpha$) and 8 equations only
(energy conservation and Eqs.~(\ref{PM1a}) to (\ref{PM2c})). The
problem can thus be solved by choosing three of the variables (say
$\omega_1$, $\vartheta_5$ and $\alpha$) as parameters. The Appendix
contains the algebraic procedure to analytically solve the problem.
In order to efficiently handle the dependence of the solutions on
the free parameters, we implemented the calculation by using the
software \textit{Mathematica} (Wolfram Research, IL).
\begin{figure}[h]
\centering
\includegraphics[width=0.4\textwidth,angle = 270]{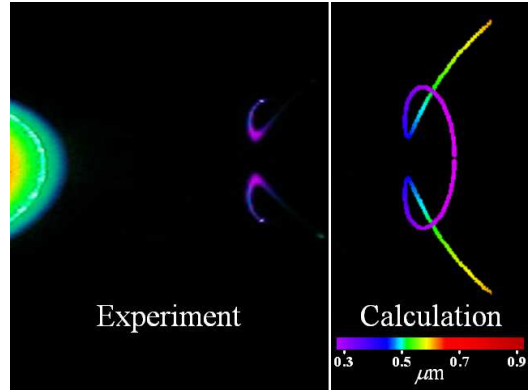}
\caption{\label{f:movie} (1.271 MB) Movie showing the variation of
the measured and calculated outputs of the crystal as a function of
the tuning angle $\alpha$ for fixed external angle between the
pumps. Left panel: picture of the output of the crystal taken with a
commercial digital photocamera (no color correction applied). LHS: a
portion of the downconversion cones. RHS: output of the upconversion
process. Right panel: calculated output of the upconversion
process.}
\end{figure}
\par
In Fig.~\ref{f:movie} we show one picture of the movie containing
the comparison between measured and calculated results for the
output of the crystal as a function of the tuning angle $\alpha$ for
a fixed value of the external angle between the two pumps. Note that
not all the calculated output frequencies appear in the experimental
part due to the sensitivity of the camera sensor. As the pictures of
the experimental outputs were taken on a screen located beyond the
crystal normally to the direction of pump $a_4$, the calculations
had to take into account both the rotation of the crystal and the
refraction of the beams at the entrance and exit faces of the
crystal. In particular, refraction at the exit face gives:
$\sin\beta_{\mathrm{j,out}} = n_{\mathrm{j}} \sin\beta_{\mathrm{j}}$
and $\sin\vartheta_{\mathrm{j,out}} =
n_{\mathrm{j}}/(1-n_{\mathrm{j}}^2
\sin^2\beta_{\mathrm{j}})^{1/2}\cos\beta_{\mathrm{j}}
\sin\vartheta_{\mathrm{j}}$.
\section{Experiment}
For the realization of the interaction described by
Eq.~(\ref{intH2}), the pump fields were provided by the
third-harmonics and the fundamental outputs of a continuous-wave
mode-locked Nd:YLF laser regeneratively amplified at the repetition
rate of 500 Hz (High Q Laser Production, Austria). In particular,
the third-harmonic field ($\lambda_4 = 349$ nm, $\sim 4.45$ ps pulse
duration) was used as the $a_4$ field producing the downconversion
cones and the fundamental field ($\lambda_5 = 1047$ nm, $\sim 7.7$
ps pulse duration) as the $a_5$ field pumping the upconversion process.\\
\begin{figure}[h]
\centering
\includegraphics[width=0.5\textwidth,angle=270]{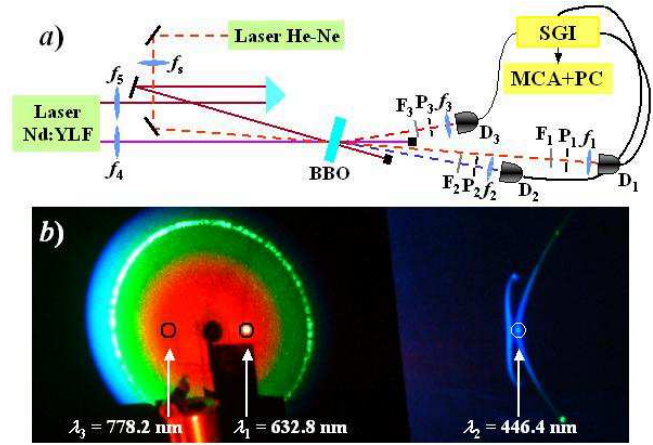}
\caption{\label{f:sper} Upper panel: scheme of the experimental
setup. BBO, nonlinear crystal; F$_{1-3}$, filters; P$_{1-3}$,
pin-holes; $f_{1-5,\mathrm{s}}$, lenses; D$_{1-3}$, p-i-n
photodiodes; SGI, synchronous gated-integrator; MCA+PC, computer
integrated multichannel analyzer. Lower panel: picture of the
visible portion of the output states on a screen located beyond the
nonlinear crystal.}
\end{figure}
As depicted in Fig.~\ref{f:sper} $a$), both pumps were focused and
injected into a $\beta$-BaB$_2$O$_4$ crystal (BBO, Fujian Castech
Crystals, China, 10 mm $\times$ 10 mm cross section, 4 mm thickness)
cut for type-I interaction ($\vartheta_{\mathrm{cut}} = 34$ deg) at
the angle $\vartheta_{5,\mathrm{ext}} = -34.8$ deg with respect to
each other. The required superposition of the two pumps in time was
obtained with a variable delay line. For alignment purposes, we used
the light emitted by a He-Ne continuous-wave laser (Melles-Griot,
CA, 5 mW max output power) to seed the process at $\lambda_1 =
632.8$ nm. The beam was collimated and sent to the BBO in the plane
containing the pumps at the external angle
$\vartheta_{1,\mathrm{ext}} = -2.54$ deg with respect to $a_4$. The
seeded interactions produced two new fields: $a_3$ ($\lambda_3 =
778.2$ nm, $\vartheta_{3,\mathrm{ext}} = 3.35$ deg) generated as the
difference-frequency of $a_4$ and $a_1$, and $a_2$ ($\lambda_2 =
446.4$ nm, $\vartheta_{2,\mathrm{ext}} = -12.78$ deg) generated as
the sum-frequency of $a_3$ and $a_5$. The pump fields were
sufficiently intense so as to observe the process starting from
vacuum fluctuations: as shown in the picture in Fig.~\ref{f:sper}
$b$), on a white screen located beyond the crystal it was possible
to see both the tunable bright downconversion cones and the two
polychromatic half-moons-shaped states generated by the upconversion
process together with the spots of the fields generated by the
seeded interactions.
\par
The conservation law in Eq.~(\ref{cm}) implies strong intensity
correlations among the generated fields which are necessary but not
sufficient to demonstrate the entangled nature of the triplet. We
realized intensity-correlation measurements by filtering each
portion of light in frequency and aligning three pin-holes of
suitable diameters on the spots of the seeded process. Note that the
seeding He-Ne light was switched-off once completed the alignment.
The light selected by the pin-holes was focused on three p-i-n
photodiodes (two, D$_{1,2}$ in Fig.~\ref{f:sper} $a$), S5973-02 and
one, D$_{3}$, S3883, Hamamatsu, Japan). The overall detection
efficiencies in the three detection arms were $\eta_1 = 0.44$,
$\eta_2 = 0.72$ and $\eta_3 = 0.43$. Each current output was
integrated by a synchronous gated-integrator (SGI in
Fig.~\ref{f:sper} $a$)) in external trigger modality, digitized by a
13-bit converter (SR250, Stanford Research Systems, CA) and recorded
in a PC-based multichannel analyzer. Data acquisition and analysis
were performed with software LabView (National Instruments, TX). We
evaluated the following correlation function on subsequent laser
shots (k)
\begin{equation}\label{Fcorr}
\Gamma(\mathrm{k})=\frac{\langle\left[m_1(\mathrm{i})-\langle
m_1\rangle\right]\left[m_{2}(\mathrm{i}+\mathrm{k})+
m_{3}(\mathrm{i}+\mathrm{k})-\langle m_{2}+m_{3}
\rangle\right]\rangle}{\sigma(m_1)\sigma(m_{2}+m_{3})}\,.
\end{equation}
in which $m_\mathrm{j}$ is the number of detected photons and
$\sigma(x)=(\langle x^2\rangle-\langle x\rangle^2)^{1/2}$ is the
standard deviation. It can be demonstrated \cite{manuscript} that in
the case of bright triplets the correlation coefficient $\varepsilon
\equiv \Gamma(0)$ must approach unity. The mean values of the
detected photons in the three parties of the triplet were $M_1 =
1.082\times 10^8$, $M_2 = 2.3\times 10^6$ and $M_3 = 1.115\times
10^8$ and the measured correlation coefficient was $\varepsilon =
0.916$.
\section{Conclusions}
The perfect agreement between calculations and experimental results
demonstrates the correctness of the developed model. This enables us
to forecast different interaction schemes that in the future will
allow to overcome the experimental difficulties experienced in the
present setup. In fact, the reason why the measured correlation
coefficient is less than unity could be the imperfect selection of
the triplet and/or the presence of spurious light in the same
location.
\section*{Acknowledgements}
This work was supported by the Italian Ministry for University
Research through the FIRB Project n. RBAU014CLC-002. The Authors
thank M.G.A. Paris (Universit\`a degli Studi, Milano) for
theoretical support on correlations.
\par\noindent
Present addresses: E. Gevinti, STMicroelectronics, via Olivetti, 2 -
20041 Agrate Brianza (MI), Italy; A. Agliati, Quanta System, via IV
Novembre, 116 - 21058 Solbiate Olona (VA), Italy.
\section*{Appendix}\label{s:app1} We describe the analytical procedure to solve the
equation system (\ref{PM1a})-(\ref{PM2c}).
\par\noindent
From the first of Eqs.~(\ref{index}) we write
\begin{eqnarray}
k_2 =
\left[G_2\cos^2\left(\vartheta_2-\alpha\right)\cos^2\beta_2+L_2\right]^{-1/2}\label{a4}\;,
\end{eqnarray}
where $G_2 = c^2/\omega_2^2\left[1/n_o^2(\omega_2)-
1/n_e^2(\omega_2)\right]= 1/k_{2,o}^2-1/k_{2,e}^2$ and $L_2 =
c^2/\omega_2^21/n_e^2(\omega_2)= 1/k_{2,e}^2$. By squaring and
summing Eqs.~(\ref{PM1a}), (\ref{PM1b}) and (\ref{PM1c}) and
defining $A = (k_4^2+k_3^2-k_1^2)/(2k_3k_4)$, we easily get
\begin{eqnarray}
\cos\beta_3 = \frac{A}{\cos\vartheta_3}\label{a9}\;,
\end{eqnarray}
whereas by squaring and summing Eqs.~(\ref{PM2a}), (\ref{PM2b}) and
(\ref{PM2c}) and Eqs.~(\ref{PM2b}) and (\ref{PM2c}) we get
\begin{eqnarray}
k_2^2 &=& k_3^2+
k_5^2+2k_3k_5\cos\beta_3\left(\sin\beta_3\sin\vartheta_5+\cos\beta_3\cos\vartheta_5\right)\label{a2'}\\
k_2^2\cos^2\beta_2 &=& k_3^2\cos^2\beta_3+
k_5^2+2k_3k_5\cos\beta_3\left(\sin\beta_3\sin\vartheta_5+\cos\beta_3\cos\vartheta_5\right)\label{a2"}\;.
\end{eqnarray}
By eliminating $k_2$ from Eqs.~(\ref{a2'}) and (\ref{a2"}) and using
Eq.~(\ref{a9}) we find
\begin{eqnarray}
\cos^2\beta_2 =\frac{k_3^2A^2\tan^2\vartheta_3+
k_5^2+2k_3k_5A\left(\tan\vartheta_3\sin\vartheta_5+\cos\vartheta_5\right)}{k_3^2+
k_5^2+2k_3k_5A\left(\tan\vartheta_3\sin\vartheta_5+\cos\vartheta_5\right)}
\label{a5}\;.
\end{eqnarray}
From Eq.~(\ref{PM1a}) we obtain
\begin{eqnarray}
\sin\beta_1 = -\frac{k_3}{k_1}\sin\beta_3\;.
\end{eqnarray}
that, once substituted into Eq.~(\ref{PM1b}) together with
Eq.~(\ref{a9}), gives
\begin{eqnarray}
\sin\vartheta_1 =-\frac{k_3 A}{\sqrt{k_1^2 -k_3^2\sin^2\beta_3}}
\label{a8}\;.
\end{eqnarray}
We now divide Eq.~(\ref{PM2b}) by Eq.~(\ref{PM2c}) and use
Eq.~(\ref{a9})
\begin{eqnarray}
\tan\vartheta_2 =\frac{k_3 A\tan\vartheta_3+ k_5\sin\vartheta_5}{k_3
A + k_5\cos\vartheta_5} \label{a6}\;.
\end{eqnarray}
By squaring Eq.~(\ref{a4}) and inserting Eq.~(\ref{a2'}) and
Eq.~(\ref{a5}) into it we get
\begin{eqnarray}
\cos^2\left(\vartheta_2-\alpha\right)=\frac{1 -L_2\left[k_3^2+
k_5^2+2k_3k_5\cos\beta_3\left(\sin\beta_3\sin\vartheta_5+
\cos\beta_3\cos\vartheta_5\right)\right]}{G_2\left[k_3^2\cos^2\beta_3+
k_5^2+2k_3k_5\cos\beta_3\left(\sin\beta_3\sin\vartheta_5+
\cos\beta_3\cos\vartheta_5\right)\right]}\label{a10}\;.
\end{eqnarray}
By substituting Eq.~(\ref{a9}) into Eq.~(\ref{a10}), expanding
$\cos^2\left(\vartheta_2-\alpha\right)$ in Eq.~(\ref{a10}) and using
Eq.~(\ref{a6}) to eliminate $\vartheta_2$, after some
simplifications we get a quadratic algebraic equation for the
variable $\tan\vartheta_3$
\begin{eqnarray}
a \tan^2\vartheta_3+b\tan\vartheta_3+c = 0\label{equ}\;,
\end{eqnarray}
whose coefficients are
\begin{eqnarray}
a &=&  G_2 A^2 k_3^2\sin^2\alpha \nonumber\\
b &=& 2 G_2 A\ k_3 (A\ k_3 + k_5 \cos\vartheta_5)\sin\alpha
\cos\alpha + 2 G_2 A\ k_3 k_5 \sin\vartheta_5\sin^2\alpha+2 L_2A\ k_3k_5\sin\vartheta_5\nonumber\\
c &=& G_2 (A\ k_3 + k_5 \cos\vartheta_5)^2
(\cos^2\alpha-\sin^2\alpha)+ 2 G_2 k_5 \sin\vartheta_5(A\ k_3  + k_5 \cos\vartheta_5)\sin\alpha \cos\alpha \nonumber\\
&+& G_2 (A^2 k_3^2  + k_5^2+2A\ k_3k_5 \cos\vartheta_5)\sin^2\alpha
+ L_2 (k_3^2 + k_5^2+2 A\ k_3k_5\cos\vartheta_5)-1
\nonumber\label{coeff}\;.
\end{eqnarray}
Once solved Eq.~(\ref{equ}), we find $\vartheta_3$ and then all the
other variables as a function of the free parameters $\omega_1$,
$\alpha$ and $\vartheta_5$.
%
\end{document}